\documentclass[conference]{IEEEtran}
\IEEEoverridecommandlockouts
\usepackage{cite}
\usepackage{amsmath,amssymb,amsfonts}
\usepackage{algorithmic}
\usepackage{graphicx}
\usepackage{textcomp}
\usepackage{xcolor}
\usepackage{tikz}
\usepackage{booktabs}

\usetikzlibrary{shapes.geometric, arrows.meta, positioning, fit, backgrounds, calc, decorations, decorations.pathreplacing}

\tikzset{
    cloud/.style={rectangle, rounded corners, minimum width=\linewidth, minimum height=0.8cm, text centered, draw=black, fill=blue!20, thick},
    edge/.style={rectangle, rounded corners, minimum width=2.5cm, minimum height=0.8cm, draw=black, align=left, fill=green!20, thick},
    runtime/.style={rectangle, rounded corners, minimum width=2.2cm, minimum height=0.6cm, text centered, draw=black, fill=orange!20, thick},
    function/.style={rectangle, rounded corners, minimum width=1.8cm, minimum height=0.6cm, text centered, draw=black, fill=yellow!30},
    arrow/.style={-Stealth, thick},
    bidiarrow/.style={Stealth-Stealth, thick},
    group/.style={rectangle, rounded corners, draw=black, dashed, thick, inner sep=10pt},
    dds/.style={rectangle, rounded corners, minimum width=\linewidth, minimum height=0.8cm, text centered, draw=black, thick}
}

\def\BibTeX{{\rm B\kern-.05em{\sc i\kern-.025em b}\kern-.08em
    T\kern-.1667em\lower.7ex\hbox{E}\kern-.125emX}}
\begin{document}

\title{A Serverless Edge-Native Data Processing Architecture for Autonomous Driving Training
}

\author{\IEEEauthorblockN{1\textsuperscript{st} Fabian Bally}
    \IEEEauthorblockA{
        \textit{Deggendorf Institute of Technology}\\
        Deggendorf, Germany \\
        fabian.bally@th-deg.de}
    \and
    \IEEEauthorblockN{2\textsuperscript{nd} Michael Schötz}
    \IEEEauthorblockA{
        \textit{Deggendorf Institute of Technology}\\
        Deggendorf, Germany \\
        michael.schoetz@th-deg.de}
    \and
    \IEEEauthorblockN{3\textsuperscript{rd} Thomas Limbrunner}
    \IEEEauthorblockA{
        \textit{Deggendorf Institute of Technology} \\
        Deggendorf, Germany \\
        thomas.limbrunner@th-deg.de}
}

\maketitle

\begin{abstract}
    Data is both the key enabler and a major bottleneck for machine learning in autonomous driving. Effective model training requires not only large quantities of sensor data but also balanced coverage that includes rare yet safety-critical scenarios. Capturing such events demands extensive driving time and efficient selection.
    This paper introduces the Lambda framework, an edge-native platform that enables on-vehicle data filtering and processing through user-defined functions. The framework provides a serverless-inspired abstraction layer that separates application logic from low-level execution concerns such as scheduling, deployment, and isolation. By adapting Function-as-a-Service (FaaS) principles to resource-constrained automotive environments, it allows developers to implement modular, event-driven filtering algorithms while maintaining compatibility with ROS 2 and existing data recording pipelines.
    We evaluate the framework on an NVIDIA Jetson Orin Nano and compare it against native ROS 2 deployments. Results show competitive performance, reduced latency and jitter, and confirm that lambda-based abstractions can support real-time data processing in embedded autonomous driving systems. The source code is available at https://github.com/LASFAS/jblambda.
\end{abstract}


\section{Introduction}

Autonomous driving systems depend on large volumes of sensor data to develop reliable perception, prediction, and planning capabilities. While extensive datasets are essential for these subsystems, increasing data volume alone does not guarantee improved performance. As shown by Makansi et al. \cite{makansi2021exposing}, the distribution of recorded scenarios, particularly the inclusion of rare yet safety critical events, plays a decisive role in determining model generalization and reliability.

Rare but safety critical events occur infrequently compared to routine driving, making their collection a major challenge for dataset design. As shown by Peng et al. \cite{peng2023pesotif}, this imbalance between common and rare scenarios significantly limits perception performance in long tail traffic situations. Capturing such events requires filtering through extensive driving data; however, performing this selection offline after recording leads to excessive storage and labeling overhead. Executing the filtering process directly on the vehicle enables early removal of irrelevant data, thereby reducing bandwidth and storage demands while preserving the diversity and relevance of the resulting dataset.

A promising approach to mitigate this imbalance is selective data recording at the edge. Instead of indiscriminate logging, the vehicle performs onboard relevance evaluation and retains only those segments deemed valuable based on criteria such as novelty, uncertainty, or scenario criticality. This strategy reduces storage and labeling costs while prioritizing events of greater significance for downstream model development. However, deploying such methods in vehicular environments introduces nontrivial challenges: algorithms must operate under strict latency, memory, and energy constraints typical of embedded systems, while accommodating the inherent heterogeneity of vehicle hardware, sensor configurations, and software stacks \cite{liu2019edge}.

To address these challenges, this work proposes a serverless edge computing framework for selective data capture. The framework abstracts platform specific concerns such as memory management and task scheduling, providing an execution environment analogous to cloud based function invocation. Developers can focus exclusively on the task specific algorithm, while the runtime manages orchestration, isolation, and deployment across vehicles. By integrating principles of Function as a Service (FaaS) computing into resource constrained automotive systems, the framework enables rapid algorithm development, simplifies integration, and enhances portability across heterogeneous embedded platforms, while maintaining compliance with the stringent latency, memory, and energy requirements inherent to on vehicle data recording.

\section{Related work}

%
\subsection{Datasets in Automotive and their Distributions}
To understand distributions in datasets, it is best to consider open datasets such as NuScenes \cite{nuscenes2019} or the Zenseact Open Dataset \cite{alibeigi2023zenseact}. Mingyu Liu \textit{et al.}~\cite{liu2024survey} provide a comprehensive study of the distributions of various open datasets across different modalities. By comparing dataset size, scenario types, citation counts, and other aspects, they derive a so-called \textit{impact score} for all datasets. Notably, the size of a dataset does not directly correlate with its \textit{impact score}.

Junyao Guo \textit{et al.}~\cite{guo2019safe} examine how scene complexity in autonomous driving datasets impacts machine learning performance. Their Driveability Assessment considers static and dynamic factors such as weather, lighting, and time of day but lacks a single quantitative metric. Although they identify diverse datasets, the study concludes that open datasets remain insufficient for robust autonomous driving algorithms. The authors propose targeted data acquisition, involving synthetic scenario creation, driver participation, and richer annotations to enhance data diversity.

\subsection{Data Filtering Algorithms for Autonomous Driving Training and Test Data}
Data engines are used to iteratively collect data that aligns with a target distribution, often leveraging machine learning models. In \cite{liang2024aide}, the authors present a data engine implementation called AIDE, which employs vision-language models (VLMs) to identify gaps in the dataset, retrain a detection model within the loop, and verify whether the retraining was successful. It is important to note that while this approach shows promising results, it is not intended to be applied to the realtime data stream of a recording vehicle.

Similarly, the work in \cite{schmidt2025joint} introduces a joint out-of-distribution filtering and active learning strategy that prioritizes informative samples during continuous data collection. The method operates within an iterative cloud-based pipeline and focuses on improving dataset diversity rather than optimizing for edge-level execution. A complementary approach is presented by \cite{lu2024activead}, where the authors propose an active learning framework for autonomous driving that selects samples based on their contribution to downstream driving performance. While these methods effectively reduce labeling effort, they assume centralized computation and abundant resources, which limits their applicability to embedded or in-vehicle environments.

\subsection{Function-as-a-Service Architectures for Edge and Automotive Applications}
Function-as-a-Service (FaaS) abstracts application logic into independently deployable and event-driven functions that execute within managed runtime environments. Li \textit{et al.}~\cite{li2022serverless} provide a comprehensive survey of serverless computing, discussing architectural components, scheduling mechanisms, and performance implications. Their work highlights that while cloud-based FaaS achieves elasticity and ease of management, it often introduces substantial overhead when applied to latency-sensitive or resource-constrained systems.

Hall \textit{et al.}~\cite{hall2019execution} present an execution model for serverless functions at the network edge, emphasizing reduced cold-start latency and efficient resource utilization through lightweight runtime management. Similarly, Shillaker and Pietzuch \cite{shillaker2020faasm} propose Faasm, a stateful FaaS runtime employing lightweight isolation to improve execution efficiency and memory sharing, which is particularly relevant for embedded workloads. Raith \textit{et al.}~\cite{raith2023serverless} conduct a recent survey on serverless edge computing, identifying key research challenges such as function placement, energy efficiency, and quality-of-service guarantees across the cloud–edge continuum. As these works address serverless execution at the edge in a broad sense, they focus on general-purpose deployment models and system-level challenges beyond the scope of this paper. In contrast, our work concentrates on a narrowly defined execution model and evaluates its impact under a controlled and consistent baseline, which motivates comparison against native implementations rather than against full-fledged serverless platforms.

In the automotive context, Alam et al. \cite{alam2023serverless} introduce Serverless Vehicular Edge Computing (SVEC), extending FaaS principles to the Internet of Vehicles. Their work focuses on offloading computation to external edge infrastructure and evaluating feasibility using emulation. In contrast, this paper targets serverless execution directly on in-vehicle hardware and evaluates runtime behavior under identical deployment assumptions. Owing to these differing execution contexts and evaluation methodologies, we provide only a high-level side-by-side positioning and do not perform a quantitative or architectural comparison.

\section{System Architecture}
The concept of lambda functions originates from cloud computing, often referred to as Serverless or Lambda Computing. A lambda function represents an abstract operation that maps input to output and executes upon a predefined condition. These functions run within a runtime framework that abstracts the underlying hardware and software, providing a modular interface for orchestrating functions independently of low level implementation details.

In contrast to cloud deployments, where lambda functions serve REST endpoints, process event streams, or perform scheduled tasks, their role in our use case differs. Here, lambda functions run directly on the recording vehicle to process sensor data and trigger recordings according to predefined conditions.

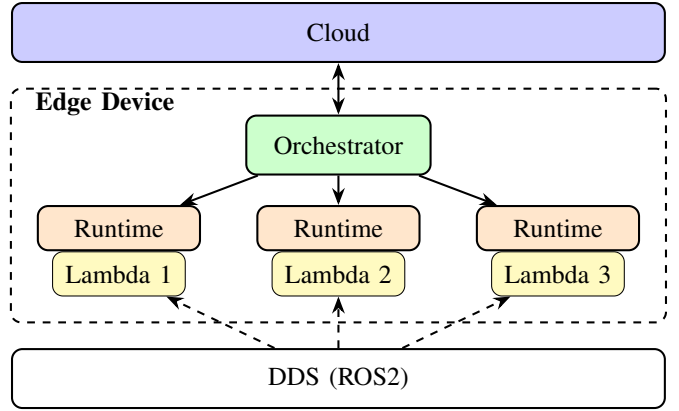
\begin{figure}[t]
    \centering
    \resizebox{\linewidth}{!}{%
        \begin{tikzpicture}[node distance=1.2cm]

            \node (cloud) [cloud] {Cloud};

            \node (orchestrator) [edge, below=0.7cm of cloud] {Orchestrator};

            \node (runtime1) [runtime, below left=0.4cm and 0.6cm of orchestrator] {Runtime};
            \node (runtime2) [runtime, below=0.4cm of orchestrator] {Runtime};
            \node (runtime3) [runtime, below right=0.4cm and 0.6cm of orchestrator] {Runtime};

            \node (lambda1) [function, below=0.0cm of runtime1] {Lambda 1};
            \node (lambda2) [function, below=0.0cm of runtime2] {Lambda 2};
            \node (lambda3) [function, below=0.0cm of runtime3] {Lambda 3};

            \node (dds) [dds, below=0.7cm of lambda2] {DDS (ROS2)};

            \draw [bidiarrow] (cloud) -- (orchestrator);
            \draw [arrow] (orchestrator) -- (runtime1);
            \draw [arrow] (orchestrator) -- (runtime2);
            \draw [arrow] (orchestrator) -- (runtime3);

            \draw [arrow, dashed] (dds) -- (lambda1);
            \draw [arrow, dashed] (dds) -- (lambda2);
            \draw [arrow, dashed] (dds) -- (lambda3);

            \begin{scope}[on background layer]
                \node [group, fit=(orchestrator)(runtime1)(runtime2)(runtime3)(lambda1)(lambda2)(lambda3), label={[yshift=-5mm, xshift=-5mm]above left:{\textbf{Edge Device}}}] (edgegroup) {};
            \end{scope}

        \end{tikzpicture}
    }
    \caption{System architecture overview. The orchestrator communicates with the cloud to fetch lambda functions and launches a dedicated runtime process for each function. Sensor data is delivered to the runtimes via DDS (ROS2), which also manages the data recorder.}
    \label{fig:architecture}
\end{figure}

\subsection{Cloud Component}
The cloud component abstracts the management of edge devices and their functions. It maintains a shared state across clients, including vehicle metadata, deployed lambda functions, and related assets, while handling repository management, deployment control, and autostart configuration. By centralizing orchestration and exposing a uniform interface, it enables coordinated deployment and updates on edge devices. This design supports reproducible experimentation and separates management from execution, allowing users to operate the system without access to low level deployment details. Authentication is provided for both users and vehicles.

\subsection{Edge Component}
The edge component is composed of two main parts: an orchestrator and the lambda runtime. Both are implemented as Rust applications, with the orchestrator launching a dedicated runtime instance for each lambda function. This distributed design simplifies implementation by delegating lifecycle management and process tracking to the operating system, eliminating the need to handle these responsibilities within a single monolithic application. The system architecture is depicted in Figure \ref{fig:architecture}.

\subsubsection{Orchestrator}
The orchestrator is responsible for managing the lifecycle of lambda functions on the edge device. It is implemented as a standalone Rust application and serves as the central coordination unit between the cloud component and the local runtime environment. Upon receiving updated function definitions or configuration changes from the cloud via a persistent control channel, the orchestrator synchronizes the set of deployed lambda functions accordingly. For each function, it instantiates a dedicated lambda runtime process, ensuring separation between different executions.

In addition to deployment and synchronization, the orchestrator collects log data and execution feedback from the individual runtimes. This information is aggregated and transmitted back to the cloud component through the same control channel, enabling monitoring, debugging, and further analysis without requiring direct access to the edge device.

\subsubsection{Runtime}
\begin{table}[t]
    \centering
    \caption{Comparisons of FaaS frameworks}
    \label{tab:comparison}
    \begin{tabular}{lccc}
        \hline
        \textbf{Framework}                  & \textbf{Data Ingress} & \textbf{RTE}  & \textbf{Function Isolation}
        \\
        \hline
        WASM V8~\cite{hall2019execution}    & REST                  & V8 / WASM     & VM                          \\
        Open Faas~\cite{alam2023serverless} & REST                  & OpenFaaS      & Container                   \\
        Ours                                & ROS2/DDS              & Python / Rust & Process                     \\
        \hline
    \end{tabular}
\end{table}
The runtime component executes individual lambda functions on the edge device. Each function runs in a separate Python process, instantiated and managed by the orchestrator. This design isolates functions from each other and provides predictable resource usage. Our process isolation approach differs from comparable FaaS frameworks as seen in Table \ref{tab:comparison}.

Lambda functions can be scheduled in two modes: \emph{periodic} or \emph{event-driven}. In periodic mode, the function executes at fixed intervals. In event-driven mode, execution is triggered by incoming data on one specific topic; in the current implementation, ROS2 provides the data distribution mechanism through topics. Restricting each lambda function to a single topic triggering execution constitutes a deliberate limitation of the current framework, which avoids arbitration between multiple concurrent trigger sources and simplifies deterministic scheduling. Multi-sensor processing is supported by sampling additional topics from their respective message queues.

The runtime exposes an API to lambda functions, which provides the following operations:
\begin{itemize}
    \item \textbf{Data access:} retrieving sensor measurements or image data from the vehicle.
    \item \textbf{Triggering:} generating actions based on computed conditions, such as starting or stopping recording.
    \item \textbf{ONNX inference:} running pre-trained machine learning models within the function.
    \item \textbf{Logging:} capturing diagnostic or application-specific output and forwarding it to the orchestrator.
\end{itemize}

\begin{figure}[h!]
    \centering
    \begin{tikzpicture}[
            >=latex, font=\small, thick,
            node distance=0.8cm and 0.8cm,
            box/.style={draw, rounded corners, fill=gray!10, minimum width=2.0cm, minimum height=0.9cm, align=center},
            smallbox/.style={draw, rounded corners, fill=gray!5, minimum width=1.7cm, minimum height=0.7cm, align=center},
            buf/.style={draw, circle, minimum height=0.55cm}
        ]

        \node[box, minimum width=2.0cm] (dds1) {DDS\\Receiver 1};
        \node[box, minimum width=2.0cm, below=of dds1] (dds2) {DDS\\Receiver 2};

        \node[buf, right=0.6cm of dds1] (rb1) {};
        \node[buf, right=0.6cm of dds2] (rb2) {};

        \node[
            draw,
            rounded corners,
            fit=(rb1)(rb2),
            inner sep=4pt,
            fill=none,
            label={[rotate=90, anchor=south, xshift=0mm, yshift=-1mm]west:\scriptsize Ring Buffers}
        ] (rbgroup) {};

        \node[draw=none, fit=(rb1)(rb2)] (rbfit) {};

        \node[smallbox, right=2.0cm of rb1, yshift=0.6cm] (start) {Start};
        \node[box, below=0.65cm of start] (runtime) {Runtime};
        \node[smallbox, below=0.65cm of runtime] (stop) {Stop};

        \draw[->, shorten >=4pt, shorten <=1pt] (dds1) -- (rb1);
        \draw[->, shorten >=4pt, shorten <=1pt] (dds2) -- (rb2);

        \draw[->, dotted] (dds1.north) .. controls +(1.0,0.5) and +(-1.0,0.4) .. (start.west) node[above,pos=0.5, font=\scriptsize]{Event Trigger};

        \draw[->] (start) -- (runtime);

        \draw[->] (runtime) -- (stop);

        \draw[->, dashed] (stop.east) to[out=30, in=-30] (start.east);

        \draw[decorate, decoration={brace, amplitude=4pt, mirror, raise=1ex}, thick]
        ($(rb2.east)$) -- ($(rb1.east)$);

        \draw[->] (runtime.west) to[out=180, in=0, looseness=1.2]
        node[above,pos=0.45, font=\scriptsize]{data access}
        ($(rbfit.east) + (0.3,0)$);

    \end{tikzpicture}
    \caption{Architecture of the runtime's concurrency model. Two DDS receivers feed ring buffers in a many-producer, single-consumer pattern. The upper topic triggers event-driven execution, while periodic execution polls both buffers.}
    \label{fig:concurrency}
\end{figure}
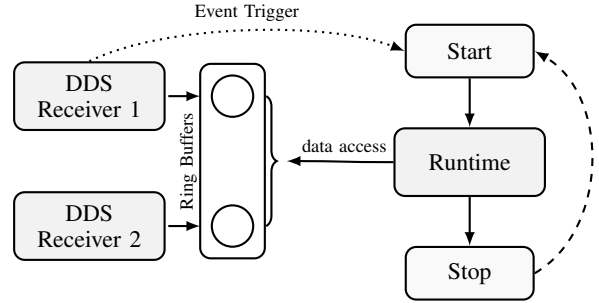

Signal data are retrieved asynchronously from the underlying Data Distribution Service (DDS) middleware. The incoming data are stored in lock-free buffers to ensure non-blocking access across concurrent threads. For high-volume data such as images, each frame is copied once upon ingress and stored in a pre-allocated memory slot managed through reference counting. This slot is then shared directly with the embedded Python runtime, enabling zero-copy data access beyond the initial transfer. The use of a fixed memory slot system guarantees memory determinism for large data streams. In contrast, low-volume data such as Inertial Measurement Unit (IMU) signals are maintained in a ring buffer without zero-copy handling. These data are re-copied into the runtime for processing, which simplifies lifetime management and eliminates the need for reference counting. In both cases, the system resembles a classical many-producer single-consumer concurrency model. Communication between the receiver threads and the runtime thread occurs either through an execution-triggering event in the case of an event-driven configuration, or without explicit synchronization when the runtime operates periodically. In both modes, the system follows a classical many-producer, single-consumer (MPSC) concurrency model, where multiple asynchronous data sources feed a single deterministic execution thread. The concurrency model is depicted in Figure \ref{fig:concurrency}.

The runtime operates independently of the orchestrator, communicating via structured messages for status updates, logging, and control commands. This separation allows the orchestrator to manage deployment and lifecycle without directly interfering with the execution of functions.

\section{Experiments}

\subsection{Setup}
All experiments were conducted on an NVIDIA Jetson Orin Nano Development Kit, an embedded system-on-chip (SoC) featuring a six-core 64-bit ARM CPU and an Ampere-based GPU with integrated Tensor Cores. The platform was chosen for its availability, low cost, and compact form factor, making it representative of resource-constrained edge devices used in real-world machine learning applications. Its heterogeneous architecture allows offloading compute-intensive inference tasks to hardware accelerators.

We compare our framework against a Python baseline across multiple lambda functions. Python is a suitable reference because it reflects common practice in edge-AI prototyping and enables a like-for-like comparison in terms of algorithmic behavior and usability. While lower-level languages such as C++ can deliver higher throughput through manual and compiler optimizations, such a baseline would primarily reflect implementation-specific optimizations rather than the impact of the framework itself.

We implement three lambda functions that continuously evaluate sensor data to decide whether it meets the criteria for recording. For comparability, identical Quality of Service (QoS) settings are used for incoming and outgoing messages, as can be seen in \ref{tab:dataset}. The evaluation focuses on camera- and IMU-based workloads; more bandwidth-intensive modalities such as LiDAR, which introduce substantially different data rates and memory characteristics, are considered outside the scope of this study and left for future work.

\subsubsection{FFT Analysis of IMU Signals}
Road surface detection helps capture data in edge cases where sensor motion is affected by vibrations from uneven terrain. Acceleration signals from the IMU are transformed into the frequency domain using a Fast Fourier Transform (FFT). Similar frequency-domain approaches have been used previously for road condition analysis, such as the work by Perttunen \textit{et al.}~\cite{perttunen2011distributed}, who applied FFT-based spectral energy analysis of accelerometer data to characterize road roughness on smartphones. Building on this concept, the spectral energy in predefined frequency bands is computed and aggregated using a weighted sum to form a \textit{Road Roughness Score}. This algorithm is used as a benchmark for CPU workloads with high-frequency signal processing requirements. We call this lambda function \textit{IMU FFT}.

\subsubsection{Multi-Sensor Evaluation}
Some driving scenarios that are worth recording require data from multiple sensor modalities to detect them. As an example, the \textit{Brake + Dark} scenario detects braking using IMU accelerometer data and estimates illumination from the mean brightness of the grayscale image captured by the front camera. This setup represents a workload combining high-frequency (accelerometer) and high-volume (camera) data, suitable for evaluating multi-sensor processing performance. We refer to this lambda function as \textit{Brake + Dark}.

\subsubsection{Object-Based Recording with YOLO11m}
Data frames containing specific object categories, such as \textit{person} or \textit{bicycle}, are considered relevant for recording. Object detection is performed using a YOLO11m model exported to ONNX and executed as an ONNX \textit{lambda} within the framework~\cite{yolo11_ultralytics}. For each input frame, the algorithm runs the detector, applies non-maximum suppression, and filters results by class and confidence threshold~$\tau$. A frame is selected for recording if at least one target object exceeds this threshold. This algorithm serves as a benchmark for evaluating the execution of neural network inference workloads on hardware-accelerated backends. This lambda function is referred to as \textit{YOLO Detector}.

\subsection{Performance Analysis}
\noindent
The performance evaluation is orchestrated by two dedicated scripts that automate all experimental components, ensuring a reproducible measurement process. To mitigate jitter, the application under test is affinitized to two isolated ARM cores of the Jetson Orin Nano, which are excluded from the operating system scheduler and freed from handling system interrupts.
\begin{table}[ht]
    \centering
    \caption{Characterization of test dataset}
    \label{tab:dataset}
    \begin{tabular}{lccccc}
        \hline
        \textbf{Signal} & \textbf{No. of frames} & \textbf{QoS}                       \\
        \hline
        OXTS IMU        & 1146                   & KeepLast(10) / Reliable / Volatile \\
        Main Camera     & 603                    & KeepLast(10) / Reliable / Volatile \\
        \hline
    \end{tabular}
\end{table}
Beyond the isolated environment, pre-recorded data from the ZOD dataset \cite{alibeigi2023zenseact} is streamed over ROS 2 DDS to the application under test. The use of openly available datasets enhances reproducibility while preserving realistic timing behavior and system load characteristics representative of live data acquisition. Involved signals from dataset alongside with a total sample number are shown in Table \ref{tab:dataset}. Upon data reception, both the lambda framework and the ROS 2 application are configured to execute a callback routine for specific message topics. Based on the functional requirements of the application under test, the incoming data is processed until a recording decision is produced.

\begin{figure}[ht]
    \centering
    \begin{tikzpicture}[
            node distance=2.5cm,
            every node/.style={font=\small},
            clock/.style={rectangle, rounded corners, draw, minimum height=1cm, minimum width=0.8cm, align=center},
            db/.style={
                    shape=cylinder,
                    draw,
                    shape border rotate=90,
                    minimum height=1.3cm,
                    minimum width=1.2cm,
                    aspect=0.25,
                    cylinder end fill=gray!30,
                    cylinder body fill=gray!20,
                    align=center
                },
            process/.style={rectangle, draw, minimum height=1cm, minimum width=2.2cm, align=center, rounded corners},
            arrow/.style={->, thick},
            decorate bracket/.style={
                    decorate,
                    decoration={brace, amplitude=5pt}
                }
        ]
        \node[db] (bag) {\shortstack{ZOD\\Dataset}};
        \node[clock, right=1.0cm of bag] (middleware) {$t_{in}$};
        \node[process, right=1.0cm of middleware] (node) {Node};
        \node[clock, right=0cm of node] (clock_out) {$t_{out}$};
        \node[db, below=0.6cm of bag] (recorder) {DDS};

        \draw[arrow] (bag) -- node[above]{DDS} (middleware);
        \draw[arrow] (middleware) -- node[above]{DDS} (node);
        \draw[arrow] (clock_out) |- node[pos=0.3, right]{RTT} (recorder);

        \draw[decorate bracket]
        ($(middleware.north) + (0,0.1)$) -- node[above=6pt]{$RTT=t_{out} - t_{in}$}
        ($(clock_out.north) + (0,0.1)$);

    \end{tikzpicture}
    \caption{Schematic of the Performance Evaluation Loop with Timestamp Alignment and RTT Measurement}
    \label{fig:performance-loop}
\end{figure}
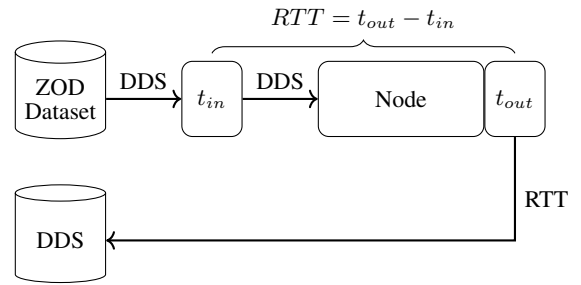
Because the application under test follows an event-driven execution model, its performance is assessed by measuring the Round Trip Time (RTT) from the moment the input data is published to the point when the recording decision is generated. This necessitates additional instrumentation in both the lambda framework and the native ROS 2 application. By transmitting supplementary RTT messages containing timestamps from both the initial data reception and the subsequent recording decision, the RTT is computed as the difference between these two recorded timestamps. Both inbound and outbound timestamps are aligned to the system clock (\textit{CLOCK\_MONOTONIC}) to ensure temporal consistency. For incoming data, a dedicated ROS 2 middleware node replaces the original timestamps in the pre-recorded dataset with the current system time before forwarding it to the application under test. Outgoing time\-stamps, on the other hand, are obtained directly through the operating system or ROS 2 API, both of which reference the same system clock. The experimental setup for measuring RTT is shown in Figure \ref{fig:performance-loop}. To minimize DDS-induced jitter between the clock alignment middleware and the application under test, the middleware node is executed within the same isolated computational environment.

For orchestrating the performance test, we loosely follow the \textit{Steady-State Performance} measuring method proposed by Crapé and Eeckhout~\cite{crape2020rigorous}. Since the lambda function under examination continuously processes sensor data in an unbounded execution loop, system startup time is excluded from the performance analysis. Accordingly, each measurement session begins with a 20-second warm-up phase during which no RTT data is collected. This is followed by three distinct 20-second execution phases, during which RTT messages are recorded. Consistent with the methodology of Crapé and Eeckhout~\cite{crape2020rigorous}, this approach ensures that transient effects such as code loading, caching, and other initialization overheads are excluded from the performance measurements.
\begin{figure}[t]
    \centering
    \includegraphics[width=\linewidth]{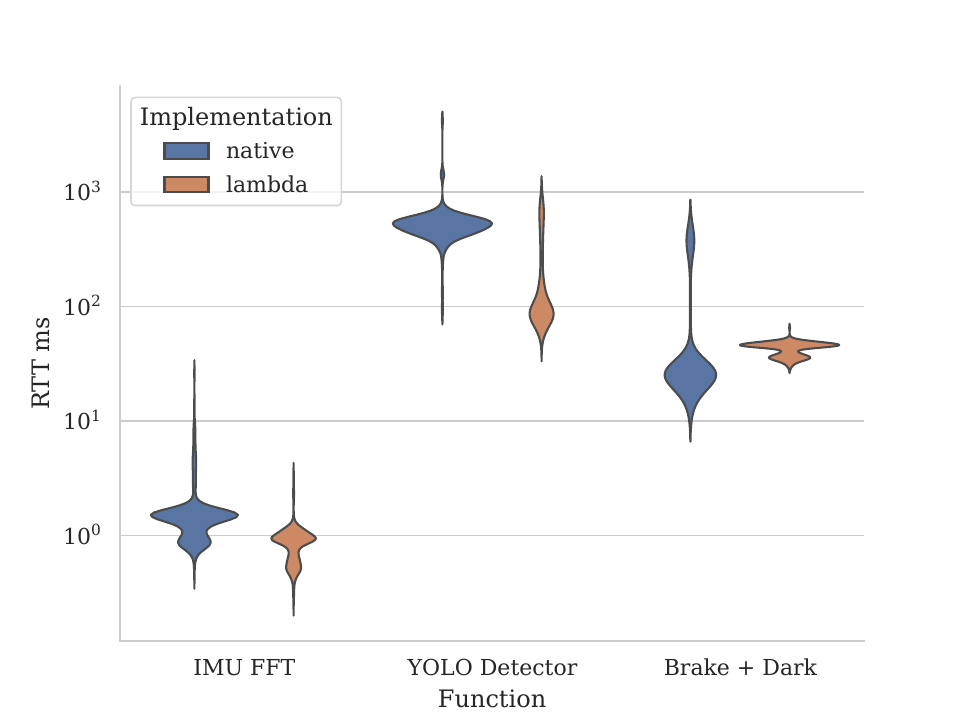}
    \caption{Results of RTT measurements for three distinct lambda functions implemented as native ROS 2 nodes and lambda functions; Lambdas demonstrate lower latency for the IMU FFT and YOLO functions and exhibit consistently reduced jitter; time scale is logarithmic.}
    \label{fig:rtt_measurements}
\end{figure}

\begin{table*}[t]
    \centering
    \caption{RTT Statistics by Function and Implementation}
    \label{tab:rtt_stats}
    \begin{tabular}{lllllll}
        \toprule
        \textbf{Function} & \textbf{Implementation} & \textbf{Min [ms]} & \textbf{Max [ms]} & \textbf{Mean [ms]} & \textbf{MAD [ms]} & \textbf{95th [ms]} \\
        \midrule
        Brake + Dark      & lambda                  & 28.43             & \textbf{65.38}    & \textbf{42.72}     & 5.06              & \textbf{49.51}     \\
        Brake + Dark      & native                  & \textbf{11.35}    & 500.06            & 71.03              & \textbf{2.94}     & 387.31             \\
        \midrule
        IMU FFT           & lambda                  & \textbf{0.24}     & \textbf{3.64}     & \textbf{0.87}      & 0.24              & \textbf{1.16}      \\
        IMU FFT           & native                  & 0.43              & 27.02             & 1.67               & 0.23              & 4.06               \\
        \midrule
        YOLO Detector     & lambda                  & \textbf{50.75}    & \textbf{898.39}   & \textbf{191.75}    & \textbf{23.08}    & 721.49             \\
        YOLO Detector     & native                  & 83.47             & 4201.67           & 532.28             & 91.43             & \textbf{663.12}    \\
        \bottomrule
    \end{tabular}
\end{table*}

\begin{figure*}[h]
    \centering
    \includegraphics[width=\linewidth]{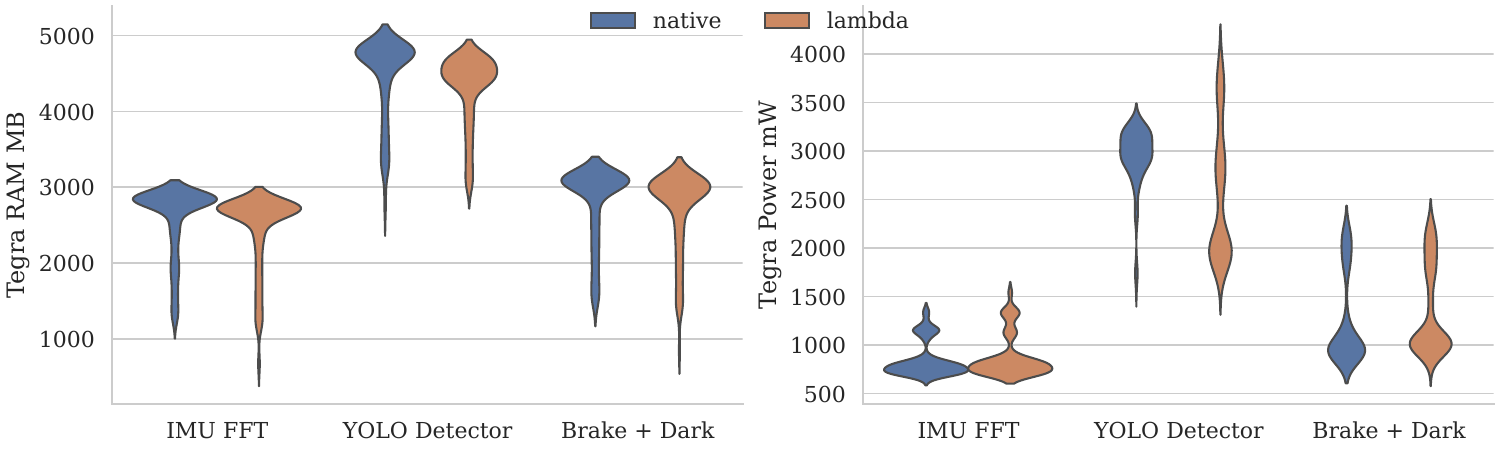}
    \caption{RAM and power consumption for the measurements displayed in figure \ref{fig:rtt_measurements}}
    \label{fig:add_measurements}
\end{figure*}

The results of the performance evaluation are presented in Fig.~\ref{fig:rtt_measurements} and summarized in Table~\ref{tab:rtt_stats}. These measurements characterize both latency and jitter across the different lambda function implementations. In terms of average Round-Trip Time (RTT), the lambda framework consistently outperforms native ROS~2 implementations, achieving approximately a twofold reduction in execution time. Regarding jitter, the IMU use case shows comparable performance between the two approaches, with both yielding a Median Absolute Deviation (MAD) of $0.24~\text{ms}$. For the YOLO detector, the lambda framework exhibits lower jitter. Conversely, in the \textit{Brake + Dark} scenario, the native ROS~2 implementation yields a lower MAD; however, a 95th percentile analysis reveals that the lambda-based execution exhibits a narrower jitter distribution overall. In Figure~\ref{fig:add_measurements}, memory and power consumption are displayed. Apart from the \textit{YOLO Detector}, both the native ROS~2 implementations and the lambda implementations show similar power and RAM profiles. For the \textit{YOLO Detector}, however, the power distribution is wider, with the main density located at lower power values compared to the native implementation.

To confirm that these improvements are not due to random fluctuations, we performed nonparametric Mann–Whitney~U tests on the RTT distributions of each workload. The differences between lambda and native implementations were highly significant ($p \ll 0.001$), indicating that the observed reductions are robust against measurement noise. We additionally controlled for systematic bias by alternating run order, equalizing core affinities, and repeating measurements under stable thermal conditions. The consistent results across all repetitions suggest that the reported latency gains stem from the framework design rather than environmental artifacts.

For the IMU use case, both implementations show similarly shaped RTT distributions. Given that the algorithm and underlying libraries are identical in both scenarios, the observed constant offset and reduced spread in the lambda framework can be attributed to its faster data ingress mechanism. In the YOLO detector use case, the performance gain observed with the lambda framework is likely attributable to the more efficient data ingress into the ONNX runtime, combined with the zero-copy transfer of image data into the Python environment, attributing to the reduction of the overhead during inference. The \textit{Brake + Dark} scenario differs notably from the other use cases, as the lambda framework exhibits higher best-case RTT values than the native ROS~2 implementation. Nevertheless, when examining the full RTT distribution, the lambda-based implementation still delivers lower average latency. We hypothesize that this behavior arises from contention in the event handling stage of the lambda framework. Specifically, the producer–consumer threading model, executing on two CPU cores, may induce competition between producer threads and the single consumer responsible for evaluating the recording condition. With two producers and one consumer, the consumer becomes the bottleneck, preventing further reduction in execution time. When the experiment is executed across all CPU cores, the best-case RTT values improve. However, without core pinning, jitter increases significantly and temporal determinism is lost.

\subsection {Limitations}
We use a black-box measurement method because we are comparing two different systems that would each need their own performance instrumentation for a white-box analysis. In our setup, the RTT is calculated using a timestamp that is added right when the application under test sends the RTT message to the DDS. This means the actual sending of the message is not included in the measurement. We consider this acceptable, since the RTT message is only sent after the main computation is finished, so the excluded time has very little impact on the overall performance results.

The data used for testing involves only two signals. Even when IMU and camera data are diverse types of sensor data found in the automotive context, the tests do not show how the system scales with more concurrent topics. While we are looking forward to also test the system with more system load, the scope of this work is to provide a proof of concept for the FaaS architecture and to proof that it is feasible to implement such as system for a small edge system as the Jetson Orin Nano. During the experiments, we observed that the compute platform was already operating near its resource limits in the YOLO detector use case, particularly with respect to available RAM.

\section{Conclusion}
With the lambda framework, we present an edge-native platform that adapts Function-as-a-Service principles to autonomous vehicle data pipelines. By abstracting task scheduling, deployment, and isolation from low-level execution concerns, the framework enables modular, event-driven data filtering across heterogeneous embedded systems, while preserving a familiar development environment inspired by cloud-native workflows.

Evaluations on an NVIDIA Jetson Orin Nano demonstrate that the framework achieves competitive performance compared to native ROS~2 deployments, reducing round-trip latency and jitter in compute-intensive and multi-sensor scenarios. These results validate the feasibility of applying lambda-based abstractions to real-time automotive data processing without compromising timing requirements. Validation on real-world in-vehicle deployments is left for future work.

In future work, the communication backbone of the framework could be extended beyond the ROS 2 DDS layer to support alternative transport mechanisms or lightweight message brokers, further improving scalability and interoperability across mixed hardware and software environments. Integrating learning-based decision modules for data relevance estimation and exploring security aspects of function isolation on shared automotive platforms represent further promising directions.

\section{Acknowledgement}
The research leading to these results is funded by the German Federal Ministry for Economic Affairs and Energy within the project “just better DATA - Effiziente und hochgenaue Datenerzeugung für KI-Anwendungen im Bereich autonomes Fahren". The authors would like to thank the consortium for the successful cooperation.

\bibliographystyle{IEEEtran}
\bibliography{ref}

\end{document}